\documentclass[aps,prl,twocolumn,superscriptaddress,raggedbottom,showpacs,floatfix]{revtex4}

\usepackage{amsmath,amsfonts,amssymb,amsthm,graphics}
\usepackage[next]{inputenc}
\usepackage[dvips]{epsfig}
\usepackage[colorlinks=true,citecolor=blue,linkcolor=blue]{hyperref}
\usepackage{bbm}
\usepackage{bbold}
\usepackage{booktabs}
\usepackage{multirow}
\usepackage{hhline}
\usepackage{amssymb}
\usepackage{comment}

\usepackage{bm}
\usepackage{color}
\usepackage{graphicx,latexsym}
\usepackage{epstopdf}

\begin{document} 
\renewcommand{\vec}{\mathbf}
\renewcommand{\Re}{\mathop{\mathrm{Re}}\nolimits}
\renewcommand{\Im}{\mathop{\mathrm{Im}}\nolimits}

%\begin{comment}

\title{Radio Frequency Tunable Oscillator Device Based on $\hbox{SmB}_6$ Micro-crystal}
\author{Alex Stern}
\thanks{These two authors contributed equally.}
\affiliation{Department of Physics and Astronomy, University of California, Irvine, Irvine, California 92697, USA.}
\author{Dmitry K. Efimkin}
\thanks{These two authors contributed equally.}
\affiliation{Joint Quantum Institute and Condensed Matter Theory Center,  University of Maryland, College Park, Maryland 20742-4111, USA}
\author{Victor Galitski}
\affiliation{Joint Quantum Institute and Condensed Matter Theory Center, University of Maryland, College Park, Maryland 20742-4111, USA}
\author{Zachary Fisk}
\affiliation{Department of Physics and Astronomy, University of California, Irvine, Irvine, California 92697, USA.}
\author{Jing Xia}
\affiliation{Department of Physics and Astronomy, University of California, Irvine, Irvine, California 92697, USA.}

\begin{abstract}
Radio frequency tunable oscillators are vital electronic components for signal generation, characterization, and processing. They are often constructed with a resonant circuit and a "negative" resistor, such as a Gunn-diode, involving complex structure and large footprints. Here we report that a piece of $\hbox{SmB}_6$, $100$ micron in size, works as a current-controlled oscillator in the $30~\hbox{MHz}$ frequency range. $\hbox{SmB}_6$ is a strongly correlated Kondo insulator that was recently found to have a robust surface state likely to be protected by the topology of its electronics structure. We exploit its non-linear dynamics, and demonstrate large AC voltage outputs with frequencies from $20~\hbox{Hz}$ to $30~\hbox{MHz}$ by adjusting a small DC bias current. The behaviors of these oscillators agree well with a theoretical model describing the thermal and electronic dynamics of coupled surface and bulk states.  With reduced crystal size we anticipate the device to work at higher frequencies, even in the THz regime. This type of oscillator might be realized in other materials with a metallic surface and a semiconducting bulk.\end{abstract}
\maketitle

\begin{figure}[t]
\label{Fig1}
\includegraphics[width=7.7 cm]{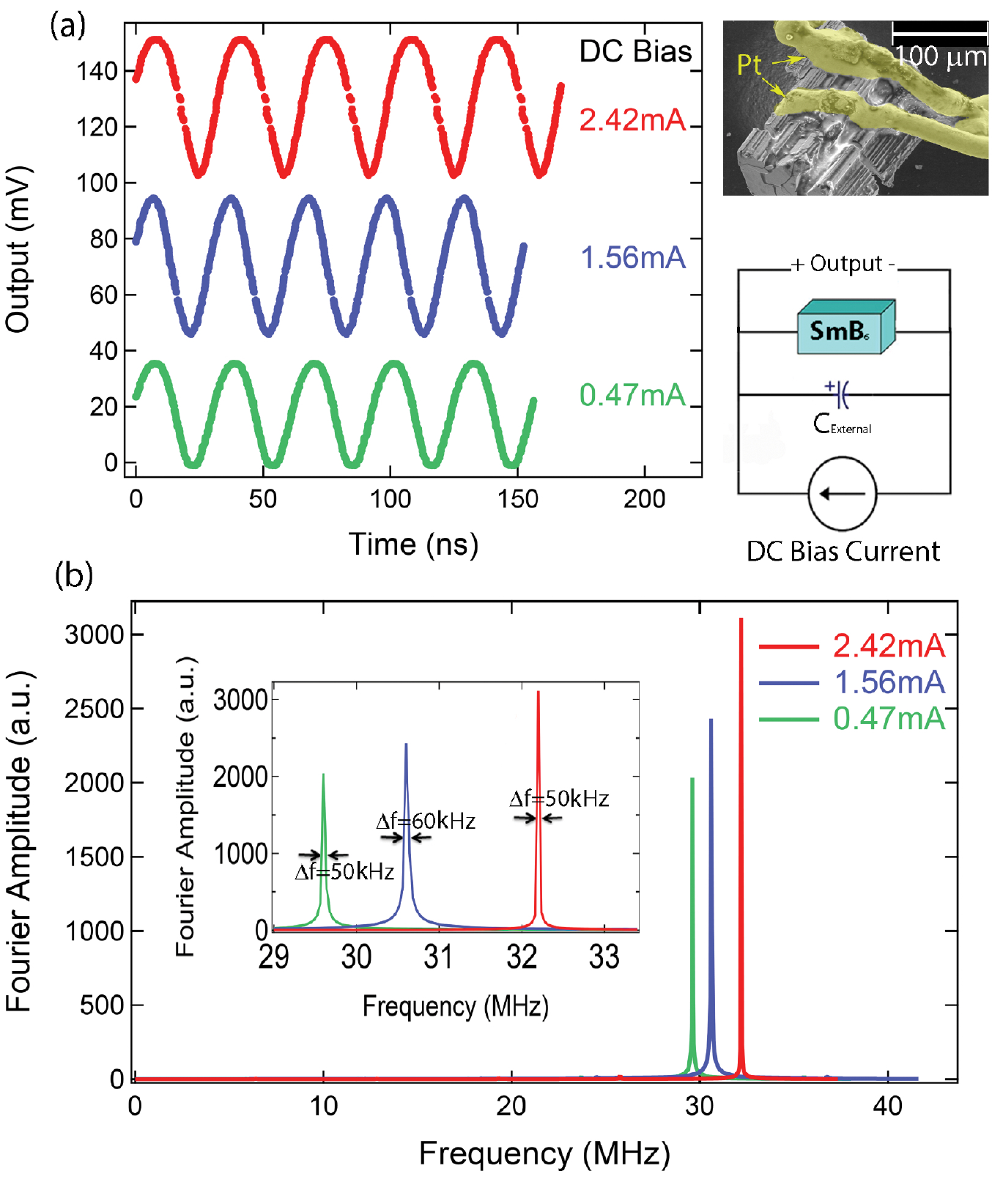}
\caption{(Color online) {\bf A $\bm{32}~\hbox{MHz}$ $\hbox{SmB}_6$ oscillator.} {\bf a}, Representative oscillation outputs at frequencies of $29.6$, $30.6$, and $32.2~\hbox{MHz}$ in ascending order, with DC bias currents of $0.47$, $1.56$ and $2.42~\hbox{mA}$.  Inset shows false-color electron microscope image of the device with platinum wires colored in yellow and $\hbox{SmB}_6$ crystal in white. The oscillator circuit consists of DC current flowing across the crystal and a capacitor (either an external capacitor or from the self-capacitance in $\hbox{SmB}_6$) in parallel. An oscilloscope is then used to measure the output waveform. {\bf b}, FFT of the data in Fig.~1-a. }
\end{figure}

Samarium Hexaboride $\hbox{SmB}_6$ is a mixed-valence Kondo insulator~\cite{Aeppli1992} with many unusual properties, perhaps most well-known for its resistance-temperature dependence that resembles both a metal and an insulator~\cite{Cooley1995}. Recently, $\hbox{SmB}_6$ was proposed to be a topological Kondo insulator~\cite{Dzero2010,Dzero2012}, with a Kondo insulating gap in the bulk and a gapless (metallic) Dirac state on the surface, which is protected by time-reversal symmetry (see Ref.~[\onlinecite{Dzero2015}] for a recent review). This metallic surface has been verified in $\hbox{SmB}_6$ by both transport~\cite{Kim2013,Wolgast2013} and ARPES~\cite{Neupane2013,Xu2013} experiments, with unusual spin texture~\cite{Xu2014}. The surface state survives non-magnetic perturbations such as electric gating~\cite{Syers2015} and mechanical abrasion~\cite{Kim2013, Wolgast2013, Syers2015}, but is destroyed by magnetic dopants that break time-reversal symmetry~\cite{Kim2014}. Quantum oscillation~\cite{Li2014} suggests that the dispersion of the surface state is Dirac-like, similar to that of graphene. The bulk of $\hbox{SmB}_6$ is extremely insulating~\cite{Kim2013, Wolgast2013} and free from impurity conductions, unless under extremely high pressure~\cite{Cooley1995} or magnetic field~\cite{Tan2015}, when the density of states of the bulk Fermi surface starts to emerge. The combination of a truly insulating bulk, a robust surface state, and strong electronic correlation~\cite{Alexandrov2015, Efimkin2014, Iaconis2015, Nikolic2014} makes $\hbox{SmB}_6$ a promising candidate to search for useful properties~\cite{Dzero2015}. Up to date, most research has been focused on the equilibrium properties of $\hbox{SmB}_6$; its transient dynamics could also be interesting.

\begin{figure}[t]
\label{Fig2}
\includegraphics[width=7.7 cm]{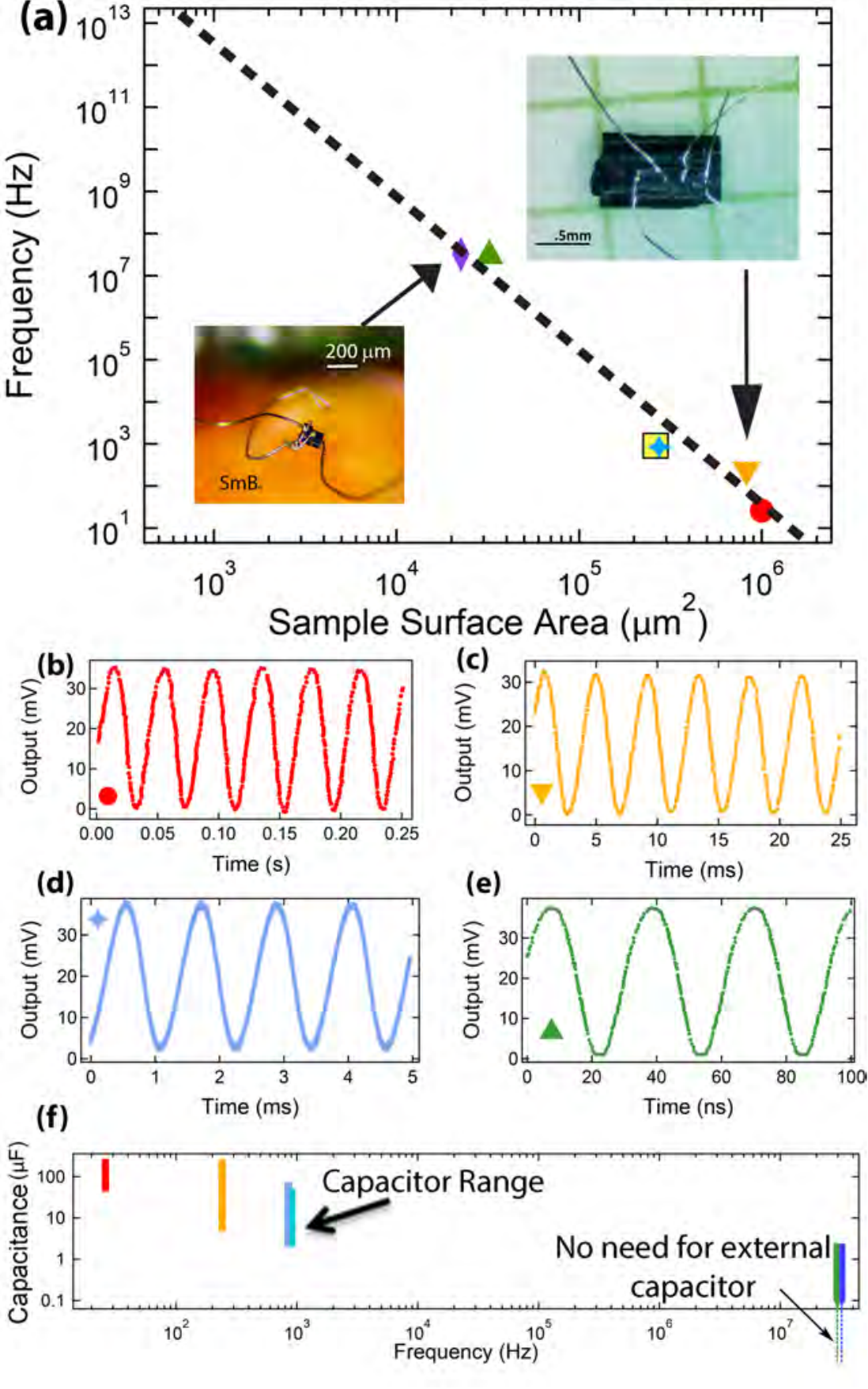}
\caption{(Color online) {\bf Scaling of frequency with crystal size.}  {\bf a}, Center frequency of oscillator devices plotted against crystal surface area.  Inserts are images of $29~\hbox{MHz}$ and $250~\hbox{Hz}$ devices. The colors of the markers on this graph match the colors in b-e. {\bf b-e}, Output waveforms of the $25~\hbox{Hz}$, $250\hbox{Hz}$, $1~\hbox{kHz}$ and $29~\hbox{MHz}$ devices respectively. The output of $32~\hbox{MHz}$ device is illustrated in Fig.~1. {\bf f}, The range of external capacitance values we used for each device. No external capacitors are needed for the two highest frequency (smallest) devices. }
\end{figure}

When biased with a few $\hbox{mA}$ of DC current, self-heating in $\hbox{SmB}_6$ causes a large nonlinear resistance~\cite{Kim2012}, which could lead to oscillation behavior. In a prior paper~\cite{Kim2012} we found that $\hbox{mm}$-sized $\hbox{SmB}_6$ crystals coupled to an external capacitor generate AC voltages at frequencies up to $\hbox{kHz}$ with a few $\hbox{mA}$ of DC bias current. While we speculated that this oscillation behavior might be related to the coupled electric and thermal dynamics of $\hbox{SmB}_6$, the exact mechanism was not understood. This was in part because the surface state had not been discovered at that time~\cite{Kim2012}, making correct modeling impossible. It was unclear how and if the intricate oscillations could be pushed to higher frequencies. With the recent advancements in the understanding of $\hbox{SmB}_6$, we are now able to develop a model with both the surface and bulk states. In this model, the origin of the oscillations lies in the strong coupling between the thermal and electrical phenomena in this surface-bulk system. Sufficient Joule heating, induced by an external DC current, can heat the bulk into a less insulating state, and trigger coupled temperature and current oscillations in both bulk and surface states. This model describes various aspects of the oscillations well and predicts that the frequency will rise sharply with reduced $\hbox{SmB}_6$ surface area, which we found to be true in $\hbox{SmB}_6$ micro-crystals. 

\begin{figure}[t]
\label{Fig3}
\includegraphics[width=7.7 cm]{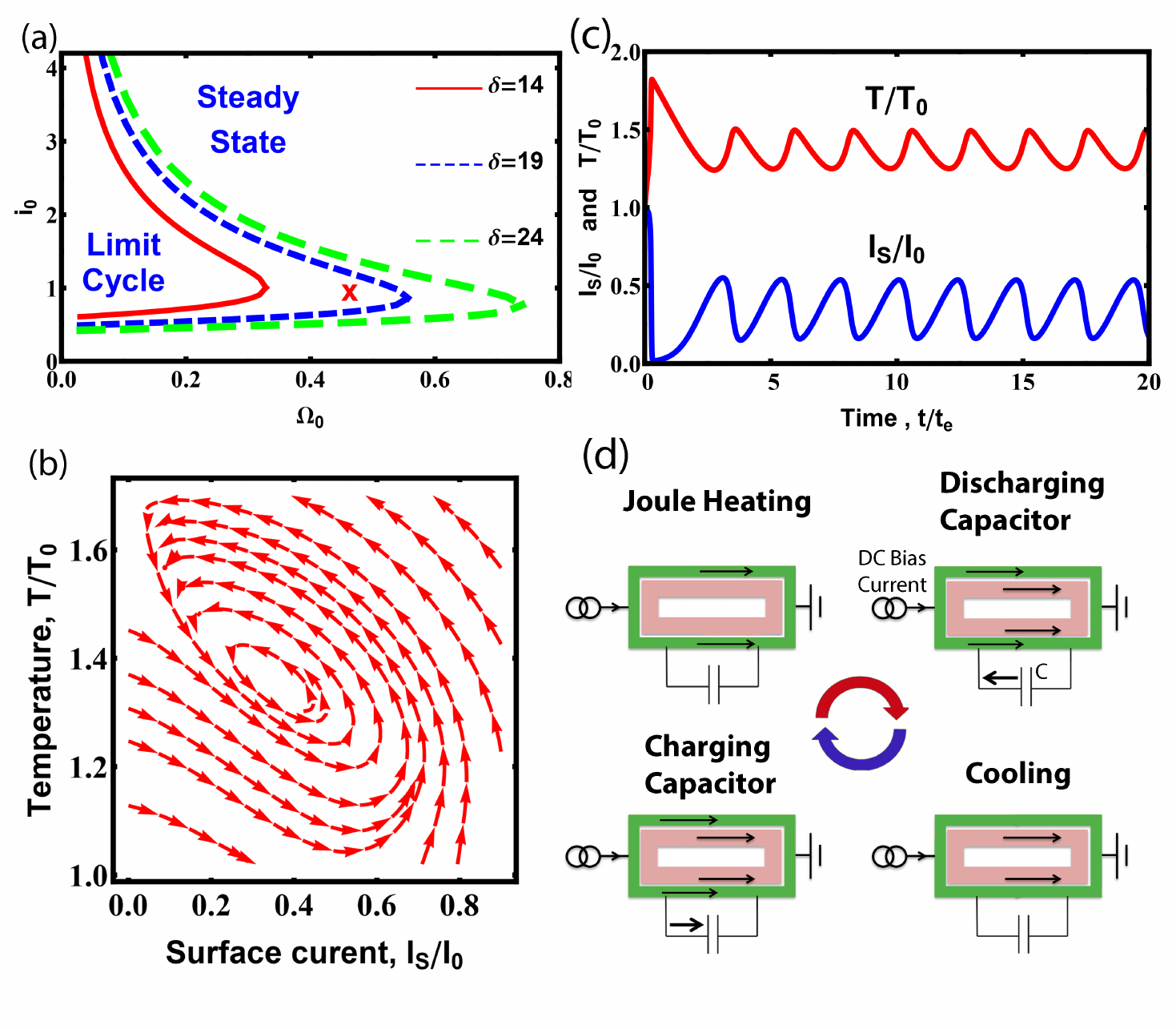}
\caption{(Color online)  { \bf Modeling the oscillation.} {\bf a}, The regime diagram of the phenomenological model. The limit cycle regime supports nonlinear oscillations of temperature and currents. The red saltire symbol denotes the parameter set ($i_0=1.0$, $\Omega=0.46$), for which time dependencies of the surface current and temperature, {\bf c}, and the corresponding vector flow plot, {\bf b}, for the model, described by the system~\ref{Model}, are presented. {\bf d}, Four phases of oscillations in order, which are described in the main text: Joule heating, discharging capacitor, cooling, and charging capacitor.  The arrows represent the flow of current.}
\end{figure}

Shown in Fig.~1 is a $31~\hbox{MHz}$ oscillator device based on a $100$-$\mathrm{\mu}\hbox{m}$-sized $\hbox{SmB}_6$ micro-crystal. DC bias currents ($I_0$) up to $4~\hbox{mA}$ are applied through two platinum wires $25~\mathrm{\mu}\hbox{m}$ in diameter that are spot-welded onto the crystal surface. And the output voltage is measured via the same platinum wires using an oscilloscope. The $\hbox{SmB}_6$ crystal itself is placed at low temperature in a cryostat, while other electronic components are held at room temperature outside the cryostat. If not explicitly stated, the measurements were performed when the cryostat is at $2~\hbox{K}$. For $\hbox{mm}$-sized $\hbox{SmB}_6$ crystals an external capacitor is required to generate oscillation~\cite{Kim2012}, such an external capacitor is found to be unnecessary for micro-crystals described here, likely due to the self-capacitance~\cite{Kim2012} we found in $\hbox{SmB}_6$. The exact origin of the self-capacitance is still unclear, but it seems to scale with the surface area, suggesting its relevance to the surface state. The oscillation behavior doesn't depend on the exact geometry of the crystal: as shown in Fig.~1 inset this oscillator is based on a rather irregularly shaped $\hbox{SmB}_6$ sample. The output of this oscillator can be continuously tuned from $29$ to $33~\hbox{MHz}$ by varying $I_0$ between $0$ and $4~\hbox{mA}$. Shown in Fig.~1-a are outputs for three representative $I_0 = 0.47$, $1.56$ and $2.42~\hbox{mA}$. And the Fourier transformations are shown in Fig.~1-b, showing a typical full width at half maximum (FWHM) spectral width $\Delta f$ of only $0.05~\hbox{MHz}$. We note that this is achieved without a phase-locked loop circuit.
 
We find that the center frequency, which is the frequency where maximum oscillation amplitude occurs, rises quickly with smaller $\hbox{SmB}_6$ crystals. Plotted in Fig.~2-a are the center frequencies for a few representative oscillators of various sizes and geometries, versus their surface areas, which we found to show the highest correlation to center frequency, compared to volume or any single dimension. Projecting the frequency-surface area scaling further, we speculate that $\hbox{THz}$ oscillations might occur for $10$-$\mathrm{\mu} \hbox{m}$-sized crystals. Operation above $2~\hbox{THz}$ is unlikely due to the $3.5~\hbox{meV}$ bulk activation gap in $\hbox{SmB}_6$. For each device, a range of external capacitors can be used to generate oscillations, as illustrated in Fig.~2-f, with no need for an external capacitor for the two highest frequency devices ($29$ and $31~\hbox{MHz}$). 

\begin{figure}[t]
\label{Fig4}
\includegraphics[width=7.7 cm]{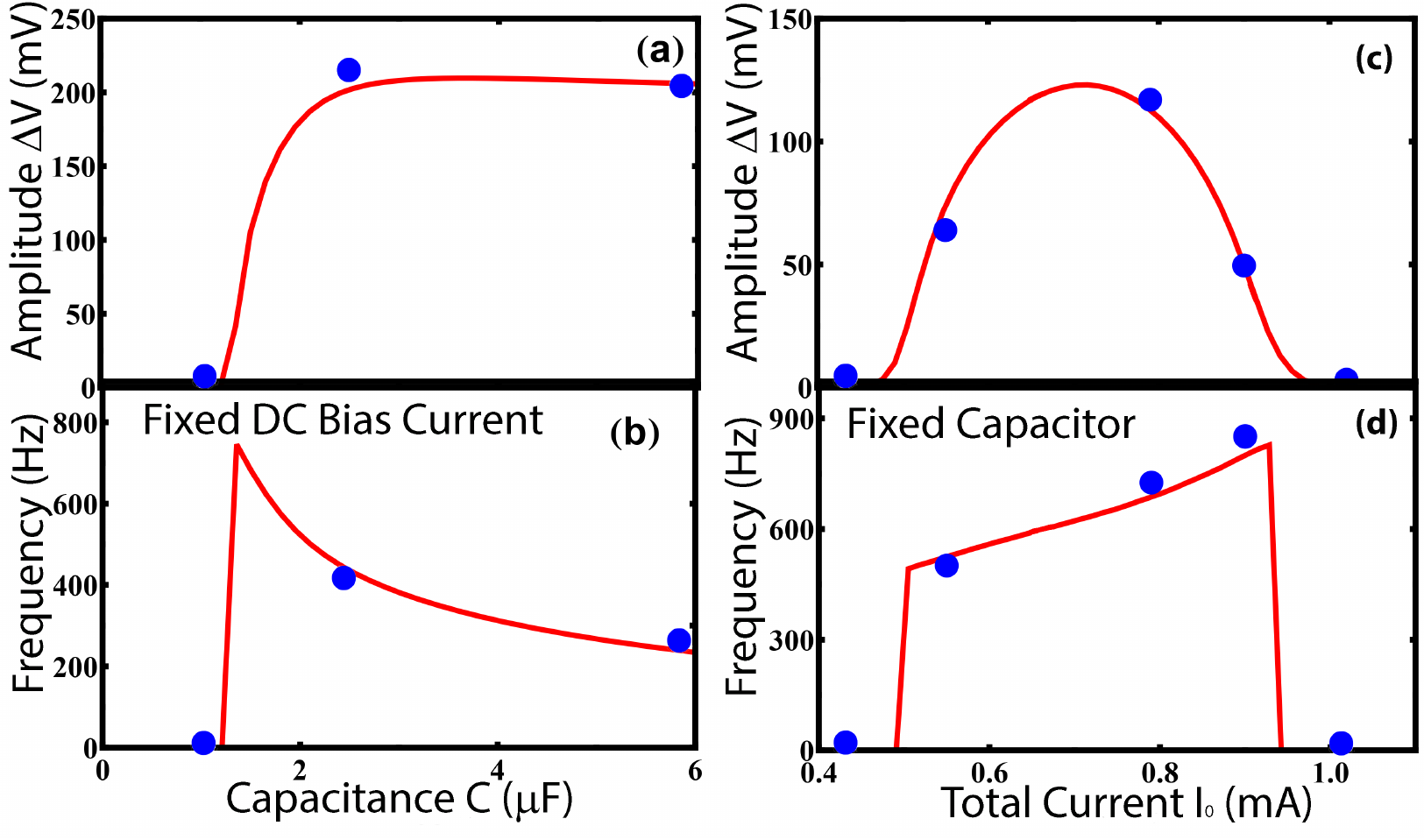}
\caption{(Color online) {\bf Comparing the model with experimental results}. Red curves are predictions from the model. Blue dots are measured valued from sample $5$. The raw experimental data is shown in supplemental Fig.~S3. {\bf a} and { \bf b}, Amplitude, $\Delta V$,  and frequency, $\nu$,  of the voltage oscillations on capacitance $C$  for fixed DC bias current, $I_0$, through the sample. Blue dotted points correspond to $I_0^\mathrm{exp}=0.55~\hbox{mA}$ , while red lines correspond to $I_0^\mathrm{fit}=0.76~\hbox{mA}$. {\bf c} and {\bf d},  Amplitude $\Delta V$,  and frequency $\nu$,  of the voltage oscillations on DC bias current $I_0$,  through the sample for fixed capacitance $C$ . The blue dots correspond to $C^\mathrm{exp}=2.2~\mu \hbox{F}$, while red lines correspond $C^\mathrm{fit}=1.5~\mu\hbox{F}$.}
\end{figure}

Both surface and bulk states are found to be essential for oscillation to occur. The oscillation amplitude diminishes at temperatures above $4~\hbox{K}$, when bulk conduction dominates; or below $1~\hbox{K}$, when surface conduction prevails. Optimal operation occurs at around $2~\hbox{K}$ when both the bulk and surface contribute to the electric conduction. This trend can be seen in the supplemental~\cite{SM} Fig.~S2-c. It is known that in $\hbox{SmB}_6$, magnetic dopants such as Gd destroy the conductive surface state~\cite{Kim2014}, while inducing little change to the bulk insulating gap. We fabricated several devices using crystals from the same $3\%$ Gd doped $\hbox{SmB}_6$ growth batch as described in~Ref.\cite{Kim2014}.  These Gd:$\hbox{SmB}_6$ samples are insulating to the lowest measurement temperature with no sign of a conductive surface. And the measured activation gap is found to be $3.3~\hbox{meV}$, which is very close to the $3.4 - 3.5~\hbox{meV}$ bulk gap in $\hbox{SmB}_6$ (supplemental~\cite{SM} Fig.~S4), suggesting its bulk is very similar to pure $\hbox{SmB}_6$. With the destruction of the surface state, no oscillation was observed from these crystals, despite testing a wide range of parameters such as temperature, bias current and external capacitance. We have also performed control experiments using another well-known Kondo insulator, $\hbox{Ce}_3$$\hbox{Bi}_4$$\hbox{Pt}_3$ (supplemental~\cite{SM} Fig.~S4, Fig.~S5) ~\cite{Hundley1990, Cooley1997}, which has a Kondo gap of $2.8~\hbox{meV}$ similar to that of $\hbox{SmB}_6$ but without a surface state at least down to our lowest measurement temperature of $1~\hbox{K}$. We couldn't find any oscillation from $\hbox{Ce}_3$$\hbox{Bi}_4$$\hbox{Pt}_3$ samples. 

Taking into account both surface and bulk states we have built a model to describe the oscillations process depicted in Fig.~3-d. It is based on charge and heat conservation equations and can be casted as follows
\begin{equation}
\begin{split}
CR_\mathrm{S} \dot{I}_\mathrm{S}=I_0-G I_\mathrm{S}; \\  C_\mathrm{H} \dot{T} =2I_\mathrm{S}^2 R_\mathrm{S} G-I_\mathrm{S} I_0 R_\mathrm{S}-\gamma(T-T_0 ).
\label{Model}
\end{split}
\end{equation}  

Here $I_\mathrm{S}$ and $I_0$ are surface and total currents through the sample; $C$ is combined internal and external capacitance;$ G=(R_\mathrm{S}+R_\mathrm{B})/R_\mathrm{B}$, where $R_S$ and $R_B=R_B^0  exp[-\Delta/T+\Delta/T_0]$ are the surface and bulk resistances with insulating gap $\Delta$; $C_\mathrm{H}=C_\mathrm{H}^0 (T/T_0 )^3$ and $\gamma$ are the heat capacity dominated by phonons and heat transfer through external leads with temperature $T_0$. $C_\mathrm{H}^0$ and $R_\mathrm{B}^0$ are heat capacity and bulk resistance in the thermal equilibrium. The detailed analysis of the model is presented in supplemental materials~\cite{SM} and here we outline our main results.                                    

In dimensionless units the dynamics of the model depend on four parameters $\rho_\mathrm{B}=R_\mathrm{B}^0/R_\mathrm{S}$, $\delta=\Delta/T_0$, $\Omega=C_\mathrm{H}^0/\gamma CR_\mathrm{S}  =t_\mathrm{H}/t_\mathrm{E}$, and $i_0=\sqrt{I_0^2 R_\mathrm{S}/\gamma T_0}$, where $t_\mathrm{H}=C_\mathrm{H}^0/\gamma$ and $t_\mathrm{E}=R_\mathrm{S} C$ are time scales of thermal and electrical processes.  The latter two can be easily tuned in the experiment by the DC bias current $I_0$ or capacitance $C$, and they control the behavior of the model. For the first two we use $\delta=18$ and $\rho_\mathrm{B}=80$, which correspond to a 0.7-mm-sized sample $5$ (supplemental~\cite{SM} Fig.~S3) at $T_0=2~ \hbox{K}$. The system of equations~(\ref{Model}) has only one fixed point (at which $\dot{I}_\mathrm{S}=0$ and $\dot{T} =0$), which is not allowed to be a saddle one. The regime diagram of the model, presented in Fig.3-a, has the steady state regime, corresponding to a stable fixed point. The regime is separated by the Hopf bifurcation line from the limit-cycle regime, supporting nonlinear time-dependent oscillations of the current and the temperature and corresponding to an unstable fixed point. Fig.~3-b and -c present the phase curves for the system (1), which illustrates the fate of the unstable fixed point, and the result of its explicit numerical integration. The oscillations, illustrated in Fig.~3-d, can be separated into four phases: 1) Joule heating. The surface current achieves maximum, while temperature is minimum. 2) Discharging of capacitor. The energy flows from electrical to thermal. 3) Cooling phase. Energy dissipates to wire leads. The current is minimized, while the temperature is maximized. 4) Charging of capacitor. The energy flows from thermal to electrical. The system is open and non-equilibrium, but during the second and fourth phases the energy of the system is approximately conserved. 

In Fig.~4 we compare the modeling results with experimentally measured oscillation behavior in sample $5$. The raw experimental data can be found in supplemental~\cite{SM} Fig.~S3. As shown in Fig.~4-a and -b, oscillations appear if the capacitance is larger than the minimal value $C^\mathrm{c}$ and only in a finite interval of currents, $I_0^{c1}<I_0<I_0^{c2}$, which corresponds to conditions $i_0^{\mathrm{c}1}<i_0<i_0^{\mathrm{c}2}$ and $\Omega<\Omega^\mathrm{c}$ in our model~(\ref{Model}) according to the regime diagram in Fig.~3-a. Illustrated in Fig.~4-c and -d, between $I_0^{\mathrm{c}1}$ and $I_0^{\mathrm{c}2}$ the dependence of the frequency on the current is linear, while the dependence of the amplitude has a bell-shaped dependence. For a fixed DC bias current, the amplitude of the oscillations increases with capacitance until saturation, while the frequency smoothly decreases. The critical values of currents and capacitance differ from sample to sample, but the behavior is general for all of them. According to our model, the oscillation frequencies are given by the inverse time scale $t_\mathrm{H}^{-1}\approx t_\mathrm{E}^{-1}$ , which drastically decreases with sample surface area, in agreement with the experimental trend (Fig.~2-a). 

While the major focus of this paper is on oscillators operating at low temperature based on proposed topological Kondo insulator $\hbox{SmB}_6$, the model developed here, in fact, describes a general system of a semiconductor and a metallic channel thermally and electrically coupled together. It is therefore in principle possible to realize such a tunable oscillator in other materials and at ambient temperatures. Candidate systems are $\hbox{Be}_2\hbox{Se}_3$/ $\hbox{Bi}_2\hbox{Te}_3$ topological insulators~\cite{Qi2011, Hasan2010}, or less-exotic semiconductor quantum well heterostructures with two-dimensional electron gas. Consider a quantum well embedded in undoped narrow-band semiconductor InAs sample with length $l\sim 1~\mathrm{\mu}\hbox{m}$, width $d\sim 1~\mathrm{\mu}\hbox{m}$ and height $h\sim 1~\mathrm{\mu}\hbox{m}$. At room temperatures $T\approx 300~\hbox{K}$ undoped InAs has resistivity $\rho_B=0.16~ \mathrm{\Omega} \hbox{sm}$, heat capacitance $C_\mathrm{H}^0=0.25~ \hbox{J}/\hbox{g K}$, heat conductivity $W=0.27~\hbox{J}/(\hbox{s sm K})$, and density $\rho_D=5.62~\hbox{g}/\hbox{sm}^3$. As a result for a typical resistance of two-dimensional electron gas $R_\mathrm{S}=500~ \Omega$, the condition $R_\mathrm{S}\approx R_\mathrm{B}$ is satisfied. The parameters of our model, given by equations~(\ref{Model}), can be estimated as $\gamma =Wld/h\approx 27 \times 10^{-6}  \hbox{J}/\hbox{sK}$ and $C_\mathrm{H}=C\rho_\mathrm{D} l d h\approx 1.3 \times 10^{-6}~\hbox{J}/\hbox{K}$. The first condition $\Omega\sim 1$ is satisfied if the time of thermal processes $t_\mathrm{H}\approx 56~ \hbox{ns}$ matches the time of electrical processes $t_\mathrm{e}=R_\mathrm{S} C$, which can be achieved for a capacitance  $C=0.1~\hbox{nF}$. The second condition, $i_0\sim 1$, is satisfied for electric current $I_0\approx\sqrt{\gamma T/R_\mathrm{S}}\approx 4~ \hbox{mA}$. The observation of oscillation in this nanostructure may demand fine-tuning of parameters; nevertheless we are optimistic that the conditions can be satisfied at room temperature. 

Acknowledgements: This material is based on research sponsored by Air Force Research Laboratory (AFRL) and the Defense Advanced Research Agency (DARPA) under agreement number FA8650-13-1-7374.

\bibliography{PRL_Oscillator}

%\end{comment}

%\begin{comment}

%%%%%%%%%% Merge with supplemental materials %%%%%%%%%%
\widetext
\clearpage
\begin{center}
\textbf{\large Supplemental Material: "Radio Frequency Tunable Oscillator Device Based on $\hbox{SmB}_6$ Micro-crystal" }
\end{center}
%%%%%%%%%% Merge with supplemental materials %%%%%%%%%%
%%%%%%%%%% Prefix a "S" to all equations, figures, tables and reset the counter %%%%%%%%%%
\setcounter{equation}{0}
\setcounter{figure}{0}
\setcounter{table}{0}
\setcounter{page}{1}
\makeatletter
\renewcommand{\theequation}{S\arabic{equation}}
\renewcommand{\thefigure}{S\arabic{figure}}
\renewcommand{\bibnumfmt}[1]{[S#1]}
\renewcommand{\citenumfont}[1]{S#1}
%%%%%%%%%% Prefix a "S" to all equations, figures, tables and reset the counter %%%%%%%%%%

\section{Device Fabrication}
We used $\hbox{Al}$ flux growth in a continuous $\hbox{Ar}$ purged vertical high temperature tube furnace with high purity elements to grow all single crystals. The samples are leached out in sodium hydroxide solution.  The surfaces of these crystals were carefully etched using an equal mixture of hydrochloric acid and water for one hour to remove possible oxide layer or aluminum residues. Samples used in the experiments were selected from a batch of samples based on their size. The exposed surfaces are $(100)$ planes. Platinum wires are attached to the samples using a micro spot welding. Measurements were carried out in a cryostat ($0.3-300~\hbox{K}$) using a standard DC voltage supply and oscilloscope.  Only the $\hbox{SmB}_6$ crystal was held at low temperature, while all other components were held at room temperature.  Low resistance wires connect the crystal to the current supply, capacitor, and oscilloscope in parallel.  

\begin{figure}[h]
\label{FigureS1}
\includegraphics[width=16 cm]{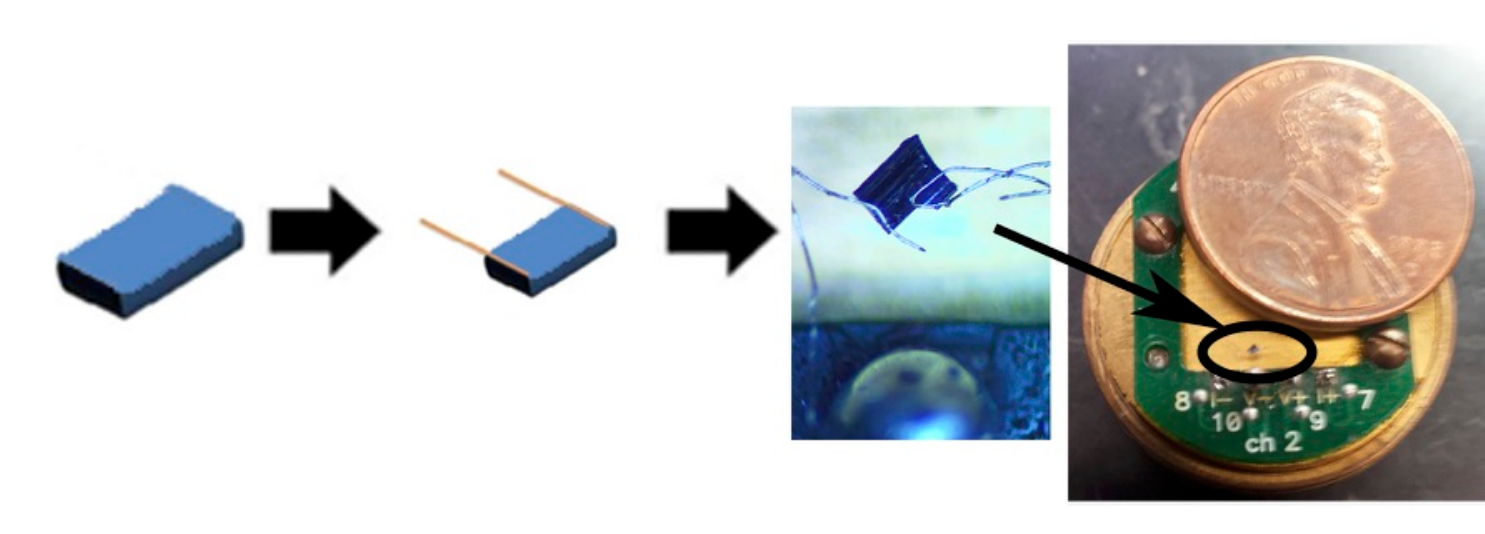}
\caption{(Color online) {\bf Device Fabrication}: The steps are described in the text in more detail.  Crystals are grown using the aluminum flux method and selected based on size with extra aluminum etched off with hydrochloric acid.  Two platinum wires are spot welded on to the sample.  Then the sample is mounted to the stage and inserted into the cryostat to be measured.}
\end{figure}

\section{A Low Frequency Oscillator}

\begin{figure}[h]
\label{FigureS2}
\includegraphics[width=16 cm]{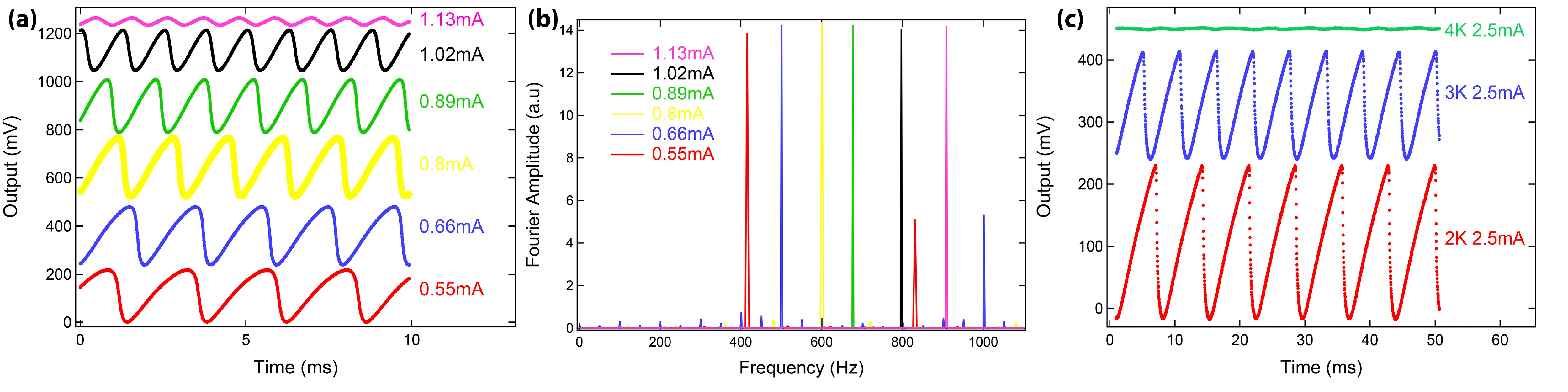}
\caption{(Color online) {\bf Low Frequency Device.} {\bf a}, The time dependence of the oscillations.  The waves become more sinusoidal and the frequency increases as the DC bias current increases. {\bf b}, The Fourier transform of the data in part {\bf a}.  The six peaks are shown to be sharp and far above other frequencies. {\bf c}, The time dependent oscillations at various temperatures.  The oscillations disappear above $4~\mathrm{K}$.  A large capacitor was used in {\bf c} than in {\bf a}, which lowers the frequency and raises the allowable current.}
\end{figure}

\section{Details of Theory Model} 

To describe voltage oscillations on the surface of $\hbox{SmB}_6$ we have developed a phenomenological model based on charge and energy conservation equations, which can be casted as follows
\begin{align}
CR_\mathrm{S} \dot{I}_\mathrm{S}=I_0-G I_\mathrm{S}; \quad \quad   C_\mathrm{H} \dot{T} =2I_\mathrm{S}^2 R_\mathrm{S} G-I_\mathrm{S} I_0 R_\mathrm{S}-\gamma(T-T_0 ).
\label{ModelSup}
\end{align}     Here $I_\mathrm{S}$ and $I_0$ are the surface and the total currents through the sample; $C$ is internal or/and external capacitance;  $G=(R_\mathrm{S}+R_\mathrm{B})/R_\mathrm{B}$, where $R_\mathrm{S}$ and $R_\mathrm{B}=R_\mathrm{B}^0  exp[-\Delta/T+\Delta/T_0]$ are the surface and bulk resistances; the former is assumed to be temperature independent, while the dependence of the latter is crucial; $C_\mathrm{H}=C_\mathrm{H}^0 (T/T_0 )^3$   is the heat capacity, which is supposed to be dominated by the bulk phonons;  $C_\mathrm{H}^0$  is the heat capacity and $R_\mathrm{B}^0$  is the bulk resistance in the thermal equilibrium at temperature $T_0$, while $\Delta$ is the energy gap in the insulating bulk;  $\gamma$ is the temperature independent heat transfer rate trough external leads, which plays the role of the bath, and $T_0$ is their temperature. The system~\ref{ModelSup} needs to be supplemented by initial conditions.  We assume that the surface current $I_\mathrm{S}=G^{-1} I_0$ settles down without any delay, while the temperature $T=T_0$ is equal to the temperature of leads. The equations for the model~(\ref{ModelSup}) are nonlinear, have reach behavior and closely capture oscillations observed in the experiment.

In dimensionless units the dynamics of the model is controlled by four parameters $\rho_\mathrm{B}=R_\mathrm{B}^0/R_\mathrm{S}$, $\delta=\Delta/T_0$, $\Omega=C_\mathrm{H}^0/\gamma C R_\mathrm{S}$, and $i_0=\sqrt{I_0^2 R_\mathrm{S}/\gamma T_0}$ ). The former two are not tuned in experiment, and weekly influence the behavior of the model. The latter two can be easily tuned by the current $I_0$ or capacitance $C$, and they control the behavior of the model. They can be rewritten as $\Omega=t_\mathrm{H}/t_\mathrm{E}$ and $i_0=I_0/I_\mathrm{Q}$, where $t_\mathrm{H}=C_\mathrm{H}^0/\gamma$ and $t_\mathrm{E}=R_\mathrm{S} C$ are the time scales of the electrical and thermal process, while $I_\mathrm{Q}=\sqrt{\gamma T_0/R_S}$ is the current sufficient to change the temperature to $\Delta T\sim T_0$. For numerical calculations we fix the first two parameters as $\delta=18$ (it corresponds to temperature $T=2~\hbox{K}$), and $\rho_\mathrm{B}=80$ (it originates from the fitting for the sample 5). It should be noted that the value of $\rho_\mathrm{B}$ cannot be directly extracted from the experimental data, since the depth of sample which temperature is affected by currents is unknown. From the value $\rho_\mathrm{B}=80$ we estimate the depth as $d\sim 2.1~\mathrm{\mu}\hbox{m}$. 
 
The system of equations~(\ref{ModelSup}) has only one fixed point (at which $\dot{I}_S=0$ and $\dot{T}=0$), which is a solution of the equation $i_0^2=G (T-T_0)/T_0$. In the vicinity of the fixed point the system~(\ref{ModelSup}) can be linearized and is given by
\begin{equation}
C R_\mathrm{S} \dot{\delta I}_\mathrm{S}=-G \delta I_\mathrm{S}-\frac{I_0 R_\mathrm{SB}}{R_\mathrm{B}} \frac{\Delta}{T^2} \delta T; \quad \quad \quad  \quad 
C_\mathrm{H} \dot{\delta T}=3 I_0 R_\mathrm{S} \delta I_\mathrm{S}+\frac{2 I_0^2 R_\mathrm{SB}^2}{R_\mathrm{B}}  \frac{\Delta}{T^2}   \delta T.
\end{equation}
with $R_\mathrm{SB}=R_\mathrm{B} R_\mathrm{S}/(R_\mathrm{S}+R_\mathrm{B})$ and the corresponding matrix we denote as $\hat{M}$. According to the general theory of dynamical systems with two degree of freedoms, the behavior of the system in the vicinity of a fixed point is defined by the signs of $\det\hat{M}$   and $\mathrm{tr}\hat{M}$. It can be shown explicitly, that $\det \hat{M}>0$ and the fixed point of the model is not allowed to be a saddle one. The value of $\mathrm{tr}\hat{M}$ can be positive (unstable spiral or nod), negative (stable spiral or nod), and the transition between these two regimes, $\mathrm{tr}\hat{M}=0$, corresponds to the Hopf bifurcation. The stable fixed point corresponds to the steady state to which the system relaxes, while from the unstable fixed point the system flows to the limit-cycle behavior, which supports nonlinear time-dependent oscillations of the current and the temperature. The phase diagram of the model is presented in Fig.~3-a for different values of the parameter $\delta$. Its area is dominated by the steady state, and only appears for a rather narrow range of parameters for the limit-cycle behavior. It appears only in an interval of currents $i_0^{\mathrm{c}1}<i_0<i_0^{\mathrm{c}2}$ and only for $\Omega<\Omega^\mathrm{c}$. The former condition demands Joule heating, induced by the currents, to be enough strong, $i_0^{\mathrm{c}1}\le i_0$, to make the bulk conductive. Nevertheless, if the system is overheated, $i_0^{\mathrm{c}2}\le i_0$, the bulk current dominates and oscillations do not appear. The latter condition implies that the time scale of thermal processes is smaller than the scale of electrical processes, which makes possible their coupled behavior. Phase boundaries $i_0^{\mathrm{c}1}, i_0^{\mathrm{c2}}, \Omega^\mathrm{c}\sim 1$ depend explicitly on parameters $\delta ,\rho_\mathrm{B}$, while the general structure of the regime diagram is insensitive to them.  The phase curves for the system~(\ref{ModelSup}), which illustrate the fate of the unstable fixed point and the appearance of the limit-cycle behavior, are depicted in Fig.~3-b. Results of the explicit numerical solution of the system in the limit-cycle regime are presented in Fig.~3-c. 

Nonlinear oscillations appear in a rather narrow range of parameters and need fine-tuning. Redistribution of current between bulk and surface is crucial and temperature oscillations should be strong enough to change the hierarchy of the surface and bulk resistances. They are supplemented by oscillations of energy between electrical and thermal states and the corresponding times should match each other. These conditions are not easy to be satisfied. The bulk of the material does not need to be topologically nontrivial, and Dirac nature of the surface states is unimportant here, nevertheless topological Kondo insulator $\mathrm{SmB}_6$ is the playground where all conditions required for the effect are naturally satisfied.

\section{Raw Data of sample 5 for Fig.~4 in the Main Text}

\begin{figure}[h]
\label{FigureS3}
\includegraphics[width=16 cm]{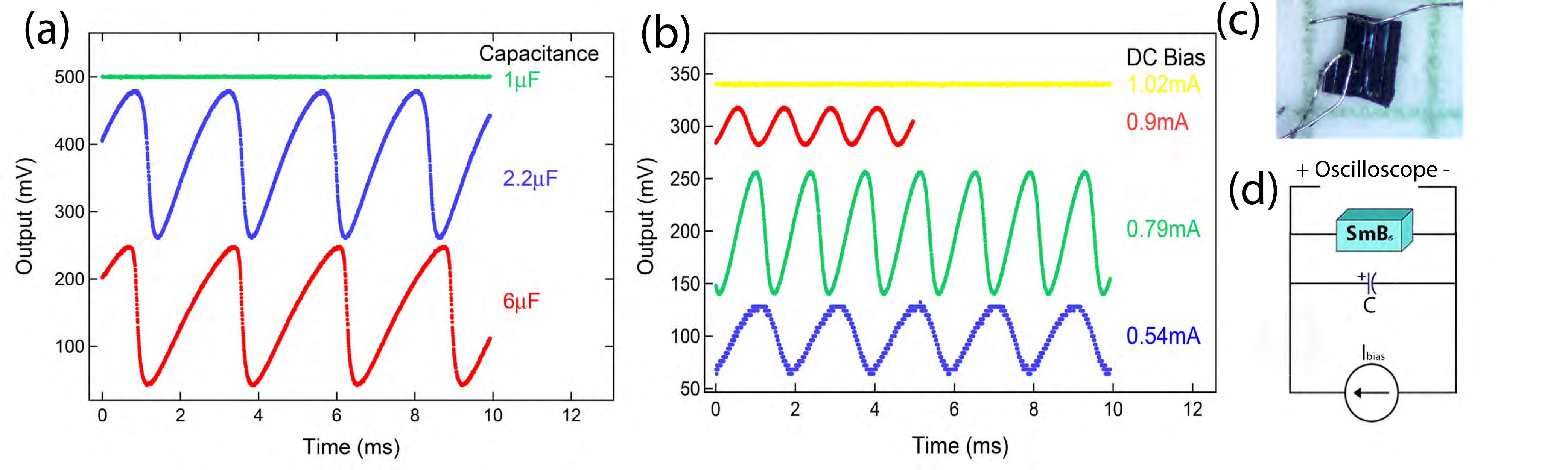}
\caption{(Color online) {\bf Sample 5, amplitude and frequency tuning via DC bias current and external capacitor matching Fig 4.} {\bf a}, The dependence of capacitance on the output.  Decreasing capacitance increases frequency until the oscillations vanish. {\bf b}, The dependence of DC bias current on the output. {\bf c}, the device used for these measurements, which is less than $1~\hbox{mm}$ in all spacial directions. {\bf d}, The schematic use to measure the oscillations.}
\end{figure}

\section{Non-oscillation in $\hbox{Gd:SmB}_6$ and $\hbox{Ce}_3$$\hbox{Bi}_4$$\hbox{Pt}_3$ devices} 

As described in the main text we have studied two sets of control samples: $\hbox{Gd:SmB}_6$ and $\hbox{Ce}_3$$\hbox{Bi}_4$$\hbox{Pt}_3$. Unlike $\hbox{SmB}_6$, no surface conduction was observed in these samples at least down to $2~\hbox{K}$. And none of them showed any oscillation despite trying a wide range of operating parameters. 

\begin{figure}[h]
\label{FigureS4}
\includegraphics[width=16 cm]{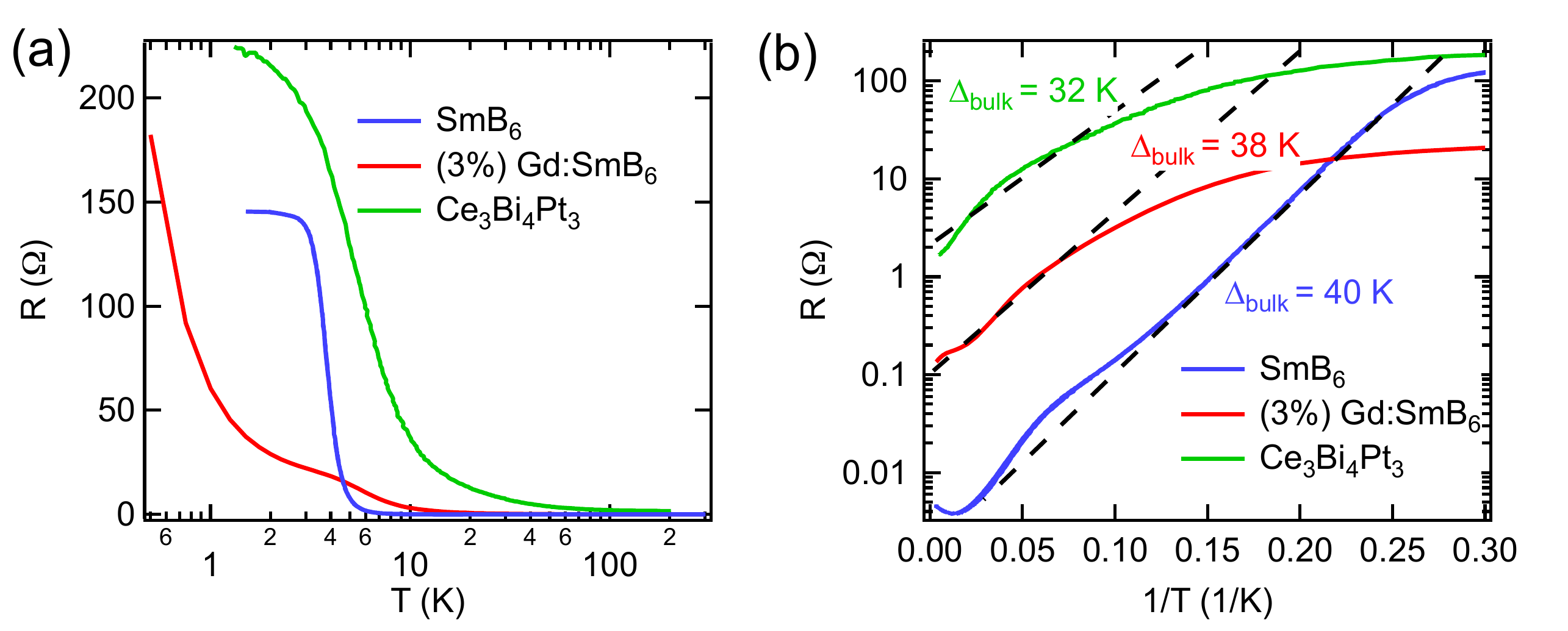}
\caption{(Color online) {\bf Transport property of $\hbox{Gd:SmB}_6$ and $\hbox{Ce}_3$$\hbox{Bi}_4$$\hbox{Pt}_3$ samples.} {\bf a}, Resistance versus temperature of 3 samples: $\hbox{SmB}_6$, $\hbox{Gd:SmB}_6$ with 3$\%$ Gd doping, and $\hbox{Ce}_3$$\hbox{Bi}_4$$\hbox{Pt}_3$. {\bf b}, Arrhenius plots of these samples in the high temperature range, showing activation gaps of $40~\hbox{K}$ ($3.5~\hbox{meV}$), $38~\hbox{K}$ ($3.3~\hbox{meV}$), $32~\hbox{K}$ ($2.8~\hbox{meV}$) for $\hbox{SmB}_6$, $\hbox{Gd:SmB}_6$, and $\hbox{Ce}_3$$\hbox{Bi}_4$$\hbox{Pt}_3$ respectively.}
\end{figure}

Shown in Fig.~S4 are the temperature dependence of resistances and the corresponding Arrhenius plots for 3 representative samples: $\hbox{SmB}_6$, $\hbox{Gd:SmB}_6$ with 3$\%$ Gd doping, and $\hbox{Ce}_3$$\hbox{Bi}_4$$\hbox{Pt}_3$. Only the $\hbox{SmB}_6$ sample shows clear resistance saturation at low temperatures due to surface state. From Arrhenius plots in the high temperature region, the activation gap of these samples are found to be rather close: $40~\hbox{K}$ ($3.5~\hbox{meV}$), $38~\hbox{K}$ ($3.3~\hbox{meV}$), $32~\hbox{K}$ ($2.8~\hbox{meV}$) respectively. A number of $\hbox{Gd:SmB}_6$ and $\hbox{Ce}_3$$\hbox{Bi}_4$$\hbox{Pt}_3$ devices, with sizes range from 0.3 mm to 1 mm have been studied at various conditions. Unlike $\hbox{SmB}_6$ none of these control samples showed any sign of oscillation. Fig.~S5 shows the voltage output of two such samples at a representative temperature of $2~\hbox{K}$ with a few representative DC drive currents. No oscillation can be discerned within experimental resolution.  

\begin{figure}[h]
\label{FigureS5}
\includegraphics[width=12 cm]{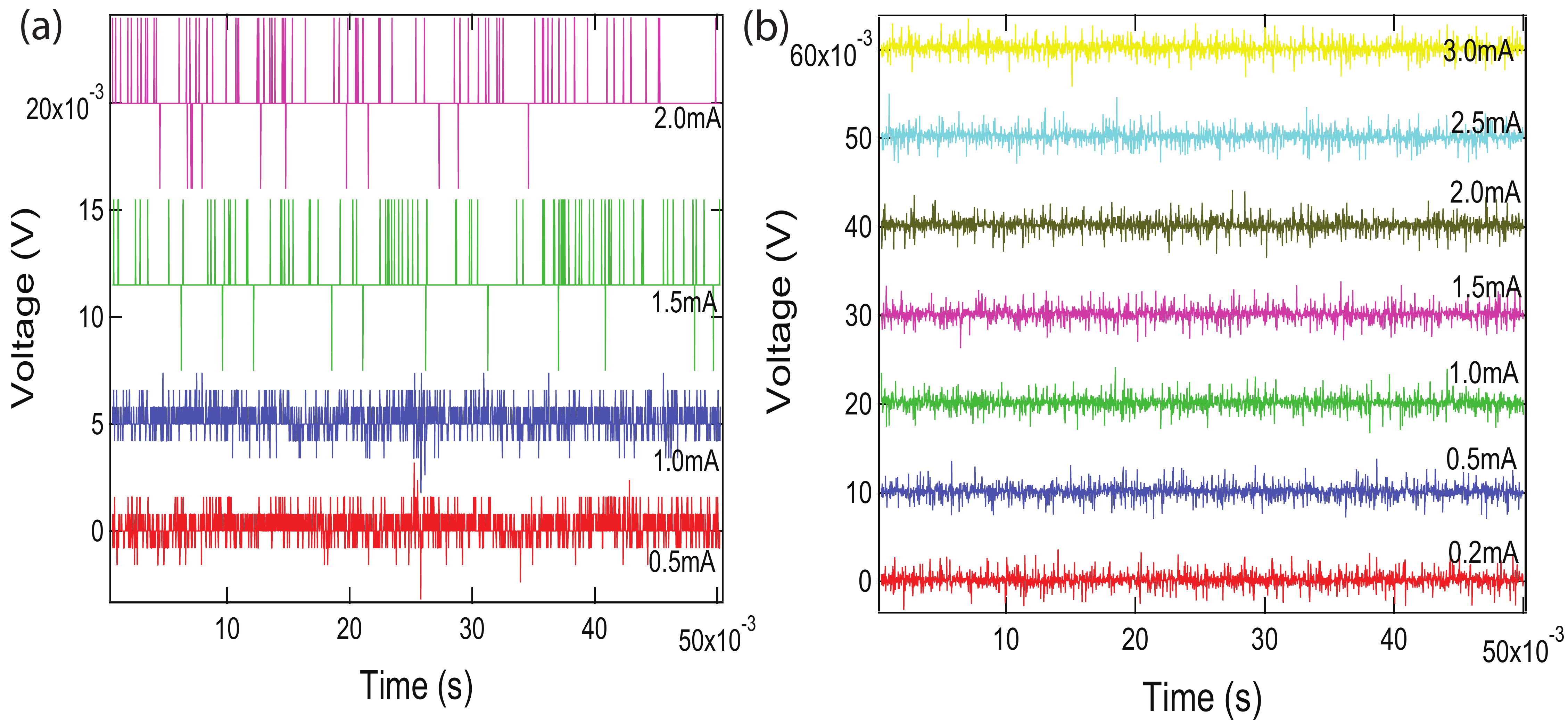}
\caption{(Color online) {\bf Non-oscillation in $\hbox{Gd:SmB}_6$ and $\hbox{Ce}_3$$\hbox{Bi}_4$$\hbox{Pt}_3$ devices.} {\bf a}, A 1 mm $\hbox{Gd:SmB}_6$ sample at $2~\hbox{K}$ with a few representative DC drive currents. {\bf b}, A 0.7 mm $\hbox{Ce}_3$$\hbox{Bi}_4$$\hbox{Pt}_3$ sample at $2~\hbox{K}$ with a few representative DC drive currents. Note that the voltage versus time curves are shifted vertically for clarity.}
\end{figure}

%\end{comment}

\end{document}